\documentclass[amsmath,amssymb, aps, pra, twocolumn,superscriptaddress]{revtex4-2}

\usepackage{float}
\usepackage{graphicx}
\usepackage{dcolumn}
\usepackage{bm}
\usepackage{hyperref}
\hypersetup{
    colorlinks=true,
    linkcolor=blue,
    filecolor=magenta,      
    urlcolor=cyan,
}
\usepackage{color}
\usepackage{fancyhdr}
\usepackage[normalem]{ulem}
\usepackage{physics}

\newcommand{\rcomm}[1]{{#1}}

\begin{document}

\title{Generation of optical potentials for ultracold atoms using a superluminescent diode}

\author{Aaron Smith}
\affiliation{School of Physics and Astronomy, University of Birmingham, Edgbaston, Birmingham, B15 2TT, UK}
\author{Thomas Easton}
\affiliation{School of Physics and Astronomy, University of Birmingham, Edgbaston, Birmingham, B15 2TT, UK}
\author{Vera Guarrera}
\affiliation{School of Physics and Astronomy, University of Birmingham, Edgbaston, Birmingham, B15 2TT, UK}
\author{Giovanni Barontini}
\email{g.barontini@bham.ac.uk}
\affiliation{School of Physics and Astronomy, University of Birmingham, Edgbaston, Birmingham, B15 2TT, UK}

\begin{abstract}
We report on the realization and characterisation of optical potentials for ultracold atoms using a superluminescent diode. The light emitted by this class of diodes is characterised by high spatial coherence but low temporal coherence. On the one hand, this implies that it follows Gaussian propagation similar to lasers, allowing for high intensities and well-collimated beams. On the other, it significantly reduces those interference effects that lead to severe distortions in imaging. By using a high-resolution optical setup, we produce patterned optical potentials with a digital micromirror device and demonstrate that the quality of the patterns produced by our superluminescent diode is consistently and substantially higher than those produced by our laser. We show that the resulting optical potentials can be used to arrange the atoms in arbitrary structures and manipulate them dynamically. Our results can open new opportunities in the fields of quantum simulations and atomtronics. 
\end{abstract}

\maketitle

\section{Introduction}
Quantum simulations with cold atoms is one of the most productive areas in quantum technology, and has provided us with important insight in condensed matter physics \cite{BlochRMP,QuantumSimRMP}, the physics of disordered systems including the observation of Anderson localisation \cite{anderson_Localization,anderson_Localization2}, low dimensional systems \cite{1DTonksGirardeau}, the use of synthetic fields \cite{synthFields} and dimensions \cite{synthDimYb}, high energy physics including Higgs modes \cite{Higgs} and black holes \cite{blackHoles}, topological effects \cite{TopologyReview} and many more. The main feature of quantum simulators based on cold atoms is the exquisite control that is nowadays possible to achieve on these systems, either by engineering the atomic internal states or by producing potentials to alter the shape, position, temperature and momentum of the samples \cite{QuantumSim}. This has been largely enabled by the rapid development of laser technology, that provides light with the ideal combination of high spatial and temporal coherence, high intensity and narrow line width. Such a combination allows one to precisely address atomic transitions with one or more photons, implement atoms traps with a large variety of shapes, produce optical lattices and dynamical potentials, and even address and manipulate single atoms \cite{QuantumSimOL,QuantumSimO2L}. 

Lately, laser light combined  with spatial light modulators is also emerging as a new avenue for the production of arbitrary and even time varying potentials \cite{Hadzibabic, PaintedPots, australian,Zupancic:16}, including non-trivial phase engineering of Bose-Einstein condensates \cite{Jones_Roberts} and applications in atomtronics \cite{ReviewStructuredLight,Atomtronics2020}. While the coherence of laser light is a key feature for the vast majority of the applications discussed, it could represent a hindrance in the generation of complex patterns using digital micromirror devices (DMDs). In these kinds of systems, the pattern produced on the DMD is imaged on the atoms and interference effects caused by the large temporal coherence can cause aberrations in the image. To mitigate this, one could use very complex and expensive optical systems or implement a feedback loop \cite{feedback}, modifying the pattern in real time on the DMD so as to minimise the difference between the projected image and the desired one. 

The methods described above attempt to mitigate the detrimental effects of temporal coherence. An alternative approach is to eradicate the problem by using light sources with reduced coherence properties. Since high spectral specifications are not required to produce optical potentials, as long as the whole spectrum of the source is far detuned, one could choose to use incoherent light sources such as LEDs to obtain pristine patterns. The main issue in this case is that the absence of spatial coherence makes it very difficult to reach intensities that allow one to produce sufficiently high optical potentials to manipulate the atomic samples. In this work, we solve this problem and produce optical potentials with significantly reduced aberrations by using super luminescent diodes (SLDs) in combination with a DMD. \rcomm{A SLD is similar to a standard laser diode, except that it strongly suppresses feedback through a combination of a tilted waveguide and anti-reflection coatings applied to the facets. This lack of feedback means that lasing is not achieved and instead the diode performs amplified spontaneous emission, with a spectrum that is not narrowed by the effects of a cavity.} SLDs are therefore temporally incoherent\cite{SLDtemporal,SLD} but feature high spatial coherence \cite{SLDHighSpatial}. We show that the low temporal coherence of SLDs allows us to produce images with significantly reduced interference effects compared to a laser. At the same time, due to its high spatial coherence, the device is capable of producing sufficiently high intensities in a well behaved Gaussian beam shape, allowing us to manipulate ultracold atomic clouds with arbitrary patterns. 

\begin{figure*}
    \centering
    \includegraphics[width=0.6\textwidth]{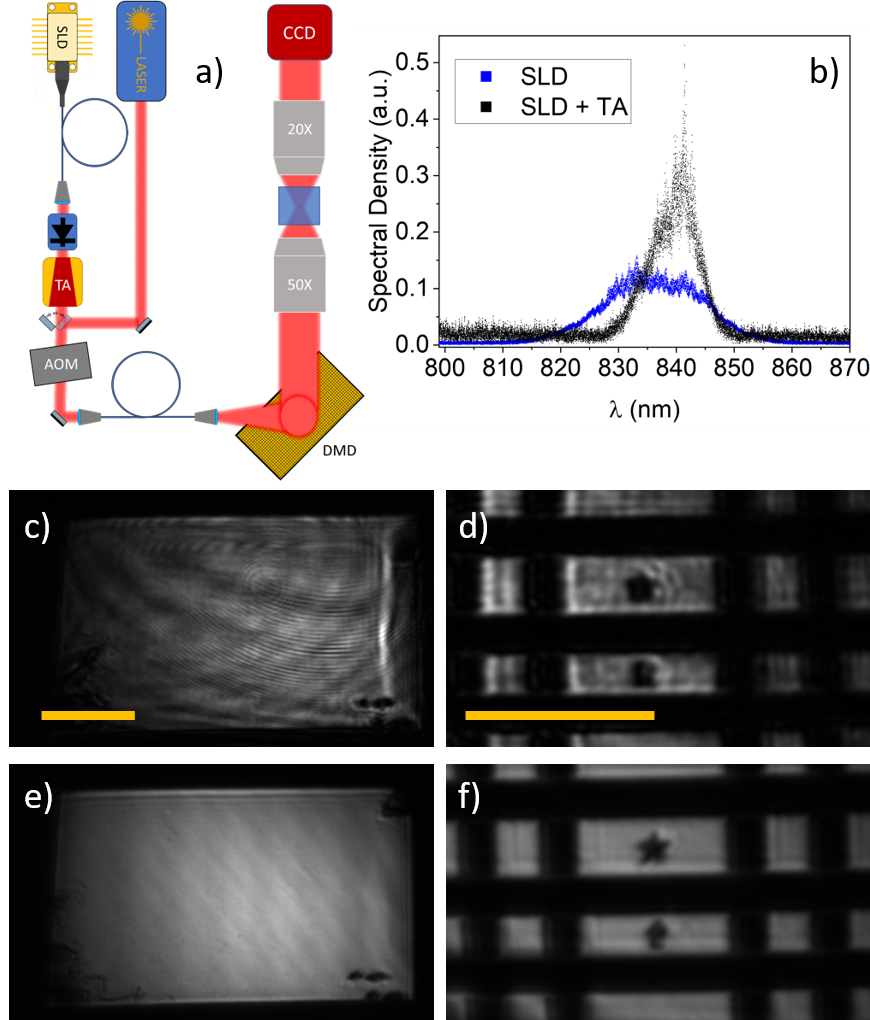}
    \caption{a) Schematic representation of our optical system. The SLD light is first amplified by a TA and then injected into an optical fibre to clean the mode. An AOM is used to control the power injected in the fibre. The output of the fibre is reflected by a DMD that imprints arbitrary patterns on the intensity profile. The patterned intensity profile is then imaged on the atoms using a 100$\times$ demagnification system that includes a 50$\times$ objective. The plane where the image is produced, that corresponds to the plane where the atoms are trapped, is then imaged on a CCD camera using a 20$\times$ magnification system. We can use the same path for the laser source by activating a flip mirror. b) \rcomm{Emission spectrum of the SLD (blue) and spectrum after the SLD emission has been amplified by a TA (black). The spectra are normalised to have the same area.} c) Image produced by the laser at the position of the atoms with the DMD fully \emph{on}. d) Same as c) but with a 2$\times$ zoom on a pattern containing an irregular mesh, a star and an arrow. Both the bars in c) and d) correspond to 50 $\mu$m. e) and f), same as c) and d) but using the SLD as light source.}
    \label{system_figure}
\end{figure*}

The paper is organised as follows: in Sec. II we describe our experimental system and discuss the characteristics of our SLD; in Sec. III we show that the SLD allows us to produce images with higher quality than with a laser, in Sec. IV we demonstrate the use of the potential created by the DMD and SLD to arrange ultracold atoms in arbitrary 1d potentials, that we also modify dynamically. In Sec. V we report our conclusions.

\section{The system}
Details about our experimental sequence and methods can be found in \cite{Jones_Roberts,DanielNJP}. In brief, we load $^{87}$Rb atoms from a 3d MOT into a crossed dipole trap and then perform evaporative cooling for 6 seconds to produce a pure Bose-Einstein condensate (BEC). For this work the final trapping frequencies are $f_r=150$ Hz and $f_x=10$ Hz, resulting in an elongated BEC with $N=2\times10^4$ atoms at $\simeq$20 nK. 

\begin{figure}
    \centering
    \includegraphics[width=0.48\textwidth]{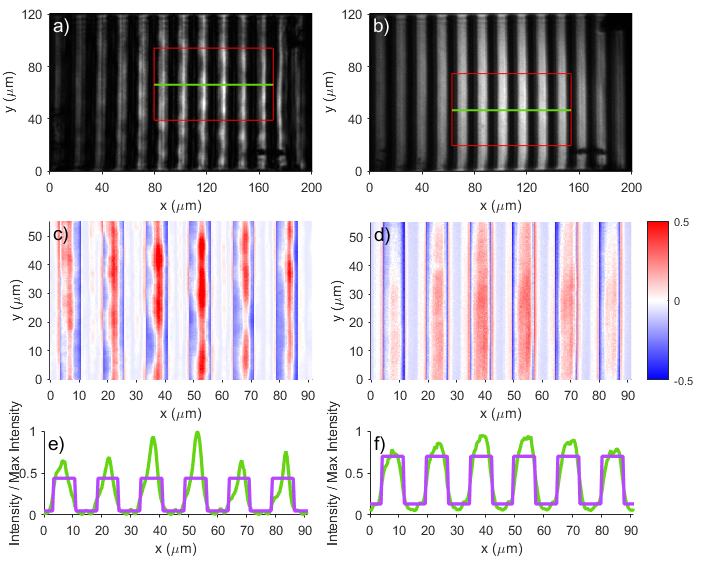}
    \caption{a) Image produced on the plane of the atoms using the laser source and a square-wave pattern on the DMD with period 1512 $\mu$m, corresponding to a period of 15.12 $\mu$m in the imaging plane. b) The same as a) but using the amplified SLD as the source. c) And d) are the residuals obtained by fitting with a square-wave grating the regions within the red rectangles in a) and b) respectively, which correspond to the beams peak intensity regions. \rcomm{The colour bar indicates the values of the residuals. The green curves in e) and f) correspond to 1d cuts along the green lines in a) and b) respectively. The magenta curves are the fits to the data using eq. (1).}}
    \label{images}
\end{figure}

To manipulate the atomic sample, we use the optical setup of Fig. 1a. Our SLD light source is a Thorlabs SLD830s-A20, \rcomm{emitting 23 mW over a spectrum with full-width at half maximum of $\Delta\lambda=$20.3 nm, centred at $\lambda_c=$836 nm, as shown in Fig. 1b. From the spectrum we estimate the coherence length of the SLD to be $l=2\ln 2\lambda_c^2/n\pi\Delta\lambda\simeq15$ $\mu$m. The light produced by the SLD is then amplified to 300 mW using a tapered amplifier (TA) with gain centered at 840 nm (Eagleyard EYP-TPA-0830).} The spectrum of the amplified light is a convolution between the gain curve of the TA and the SLD spectrum. This is shown in Fig. 1b, where it is possible to appreciate that after the amplification the spectrum is centered at $\lambda_c'=$840 nm and its full-width at half maximum is $\Delta\lambda'=$9.4 nm, therefore smaller than that of the bare SLD. \rcomm{This results in a doubly longer coherence length of $l'\simeq$ 33 $\mu$m.} Care must be taken to provide 30 dB of isolation between the TA and the SLD to prevent  damage to the latter with the TA back emission. The amplified light passes through an acousto-optic modulator that allows us to control its power and then is injected into a single mode optical fibre. A flip mirror allows us to use the same acousto-optic modulator and fibre with a laser tuned at 800 nm, enabling a direct comparison between the two light sources \footnote{\rcomm{We do not observe any difference in the acousto-optic modulator efficiency between the two light sources.}}. The light coming out from the optical fibre is reflected by a DMD and then sent onto the atoms along the vertical direction, demagnified by a factor 100 using an optical setup that includes a 50$\times$ high-resolution optical microscope objective. The DMD has $1920\times1080$ micromirrors of size 10.8 $\mu$m. Each micromirror can be individually set \emph{on} or \emph{off} every 100 $\mu$s. After reflection by the DMD and transmission through the optics, we typically obtain 1 mW of power at the position of the atoms \footnote{\rcomm{We note that the reflection from our DMD is only $\simeq$10\%, although we expected $\simeq$50 \% reflectivity according to the specifications of the device.}}.

\begin{figure}
    \centering
    \includegraphics[width=0.48\textwidth]{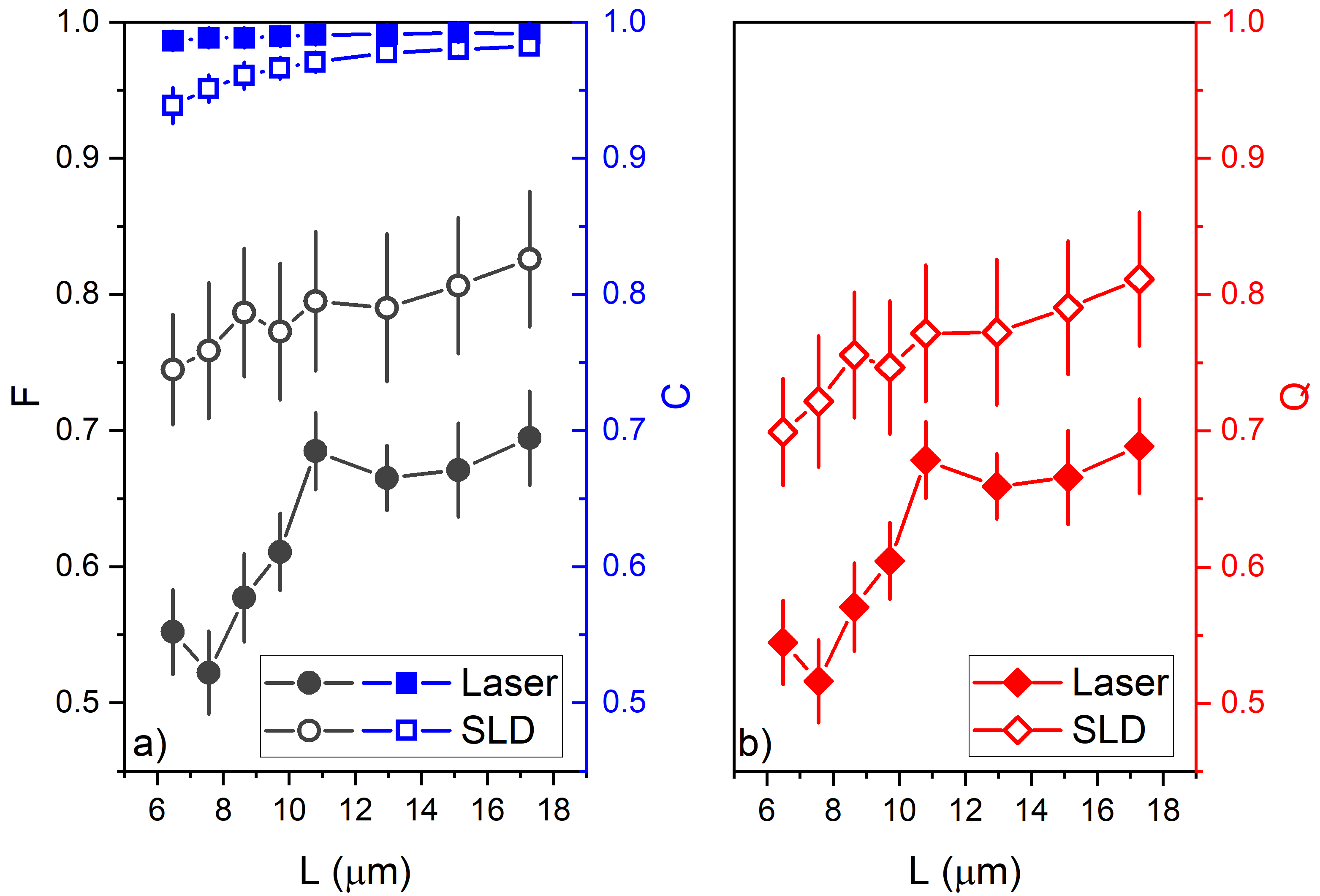}
    \caption{a) Fidelity (circles) and contrast (squares) of the SLD images (open symbols) and laser images (filled symbols) as a function of the semi-periodicity of the square wave $L$. b) Quality of the image as a function of $L$ for the SLD images (open symbols) and the laser images (filled symbols).}
    \label{figFCQ}
\end{figure}

The images produced by our system at the position of the atoms are imaged on a CCD mounted in the vertical direction, using a 20$\times$ magnifying system, resulting in a resolution of $\simeq2$ $\mu$m. The same imaging system is used to perform high-resolution absorption imaging of the atoms. In Fig 1c-e we show two examples of the images obtained using each of the two light sources. The SLD-generated images presents sensibly reduced interference fringes, are overall more uniform and feature finer details. 

\section{Image characterisation}
In this section we quantitatively compare the images produced using the SLD and the laser. For simplicity and clarity, we constrain our analysis to one-dimensional structures, as our findings can be easily extended to two dimensions. In particular, we investigate the ability of the two sources to produce square-wave gratings with our optical system, as in the examples in Fig. \ref{images}a and b. In order to quantitatively compare the images produced by the two sources, we use two parameters: the \emph{fidelity} and the \emph{contrast}. \rcomm{We have chosen to separately evaluate these two parameters to highlight the superior ability of SLDs to produce square-wave patterns, regardless of their amplitude or residual background}. To compute the fidelity, we perform two-dimensional fits of the images produced on the CCD $I$ with the ``target image" \rcomm{$I_0(i,j)=A+B\vartheta[\cos(\pi id/L+\phi)]$, i.e., a perfect square-wave grating, with $\{i,j\}$ the pixel coordinates in the two directions, $d$ the pixel size and $L$ the semi-period of the square-wave. $A, B$ and $\phi$ are free fitting parameters}. The fidelity is then defined as:    
\begin{equation}
    F=1-\frac{1}{BN}\sqrt{\sum_{i,j}(I(i,j)-I_0(i,j))^2}
\end{equation}
where $N$ is the number of CCD pixels in the region of interest. As an example, in Fig. \ref{images}c and d we report the residuals $I(i,j)-I_0(i,j)$ obtained by fitting the outlined regions in Fig. \ref{images}a and b with the same square-wave function. We consistently observe that for SLD-generated images the values of the residuals are lower all across the images, and also that the images have fewer aberrations. \rcomm{This is also evident in Fig. \ref{images}e and f, where we show the 1d cuts across the green lines in Fig. \ref{images}a and b together with the fits done using eq. (1).} To evaluate the contrast, we integrate the images along the $y$ direction obtaining one-dimensional arrays $\tilde{I}$  and then compute
\begin{equation}
    C=\frac{\text{max}(\tilde{I})-\text{min}(\tilde{I})}{\text{max}(\tilde{I})}.
\end{equation}

The fidelity quantitatively estimates how close the image produced is to the target image, while the contrast accounts for the ability of a certain source to effectively produce dark regions, \rcomm{indeed it varies linearly with $\text{min}(\tilde{I})$}. For the implementation of accurate optical potentials for cold atoms we require both these quantities to be as close to 1 as possible. Indeed, low values of $F$ indicate aberrations in the optical potential and therefore a limited capacity to accurately manipulate the atoms. Low values of $C$ imply instead that higher intensities need to be used to reach the desired potential depth, causing an increase in the scattering rate that makes the potential increasingly less conservative. To account for both these effects we use the \emph{quality of the image} $Q=F\times C$ as our figure of merit. An image with $Q=1$ would perfectly produce the desired optical potential.

In Fig. \ref{figFCQ} we report the behaviour of $F$, $C$ and $Q$ as a function of $L$. With $L$ we indicate the semi-periodicity in the imaging plane, therefore $L$ is a factor 100 smaller than the semi-periodicity of the pattern on the DMD. The SLD images are characterised by a slightly lower contrast than the laser images, especially for low values of $L$. \rcomm{Our measurements therefore confirm that temporal coherence plays a minor role in the contrast of an image \cite{coherence}, but also suggest that as the size of the features is reduced, such role becomes increasingly important}. Notably, SLD images consistently feature a substantially higher fidelity, as the reduced temporal coherence of the SLD almost completely eliminates interference effects. The main source of loss of fidelity for the SLD images is the underlying Gaussian distribution of the intensity, as can be seen for example in Fig. 1e and 2b,d. The overall effect is that the SLD generates images with values of $Q$ higher by at least 0.1 than those generated by the laser over the whole range explored. Our imaging system does not allow us to perform quantitative measurements for $L< 5$ $\mu$m. \rcomm{To provide a qualitative estimate below the limits of our imaging system, we fit $C$ and $F$ with second order polynomial functions. By doing this we infer that the SLD images should feature a higher $Q$ down to sub-$\mu$m structures.} 

\begin{figure}
    \centering
    \includegraphics[width=0.48\textwidth]{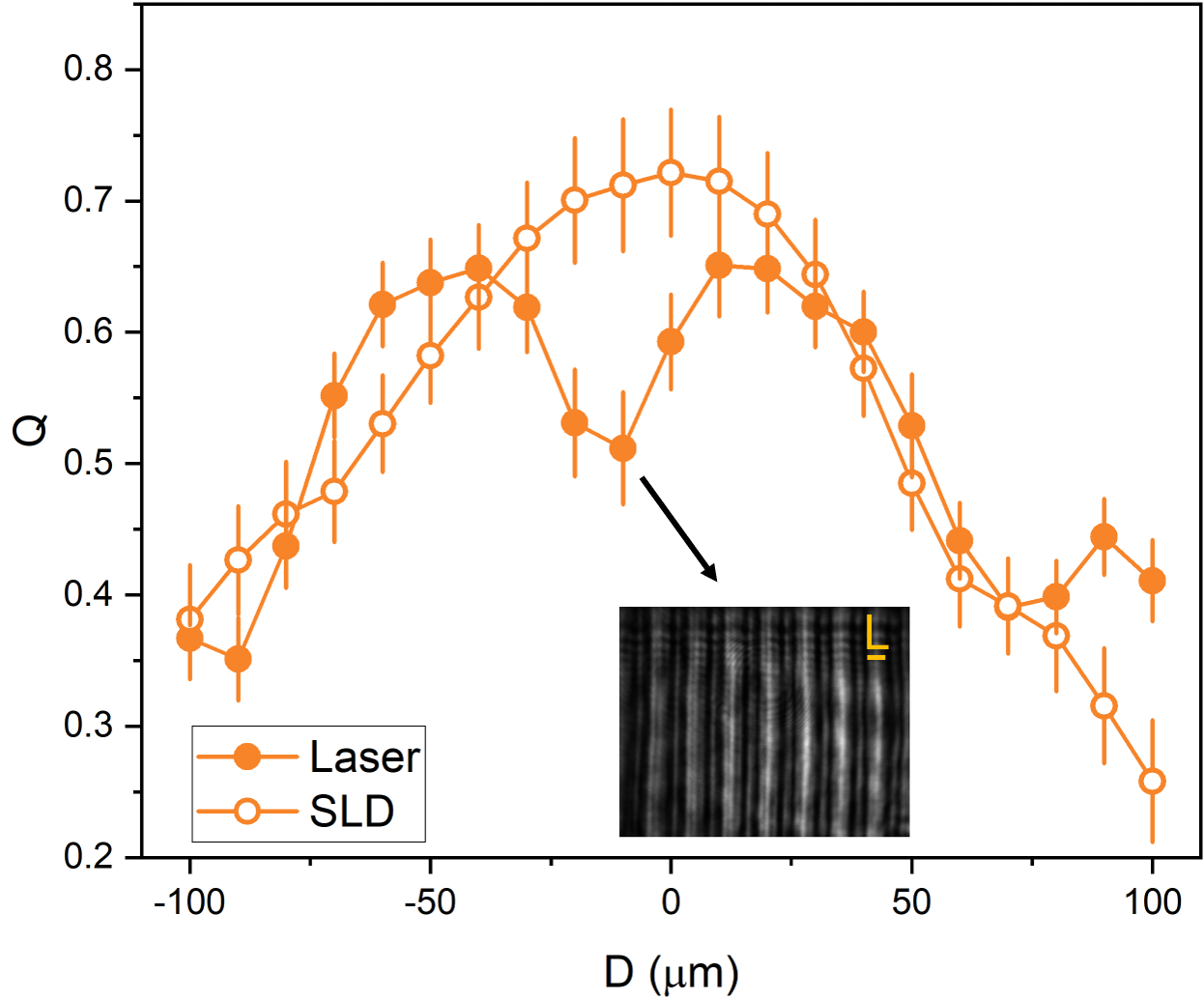}
    \caption{Quality of the image generated by the SLD (open symbols) and the laser (filled symbols) for $L$=7.56 $\mu$m a function of the distance from the focal plane $D$. For SLD-generated images $Q$ is a smooth function of $D$, while for laser-generated images $Q$ is modulated by the Talbot effect, that every $\simeq$80$\mu$m generates images with semi-periodicity $L/2$, as shown in the inset. }
    \label{figfocus}
\end{figure}

Another important feature stemming from the partial incoherence of SLD-generated light is that it is free from the Talbot effect, that alters the periodicity of regular structures as the beam propagates. This can be observed in Fig. \ref{figfocus}, where we report the quality of the image $Q$ as a function of the distance from the focal plane $D$ for $L=7.56$ $\mu$m. We observe that for SLD-generated images $Q$ is a smooth function of $D$, while for laser-generated images $Q$ presents some local maxima and minima. The minima are due to the Talbot effect, \rcomm{that at integer fractions of the Talbot length $\Lambda=2(2L)^2/\lambda$} causes the appearance of images with $L/2$ semi-periodicity. This in turn causes a degradation both in $F$ and $C$. SLD-generated images feature also higher values of $Q$ over a broad range of $D$. The fact that the images produced by the SLD are free from the Talbot effect can be relevant for the implementation of high-quality 3d-patterns, lattices and holograms using two perpendicular beams that carry periodic or near-periodic structures. 

\section{SLD-generated optical potentials for cold atoms}
Besides the low temporal coherence that enables the generation of optical potentials with higher quality than lasers, SLDs feature high spatial coherence, that in turn implies Gaussian propagation. On the one hand this represents a problem in reaching very high values of $Q$, as demonstrated in the previous section. On the other, this is a crucial feature to reach sufficiently high intensities and generate optical potentials that can be used to manipulate ultracold atoms.   

\begin{figure}
    \centering
    \includegraphics[width=0.48\textwidth]{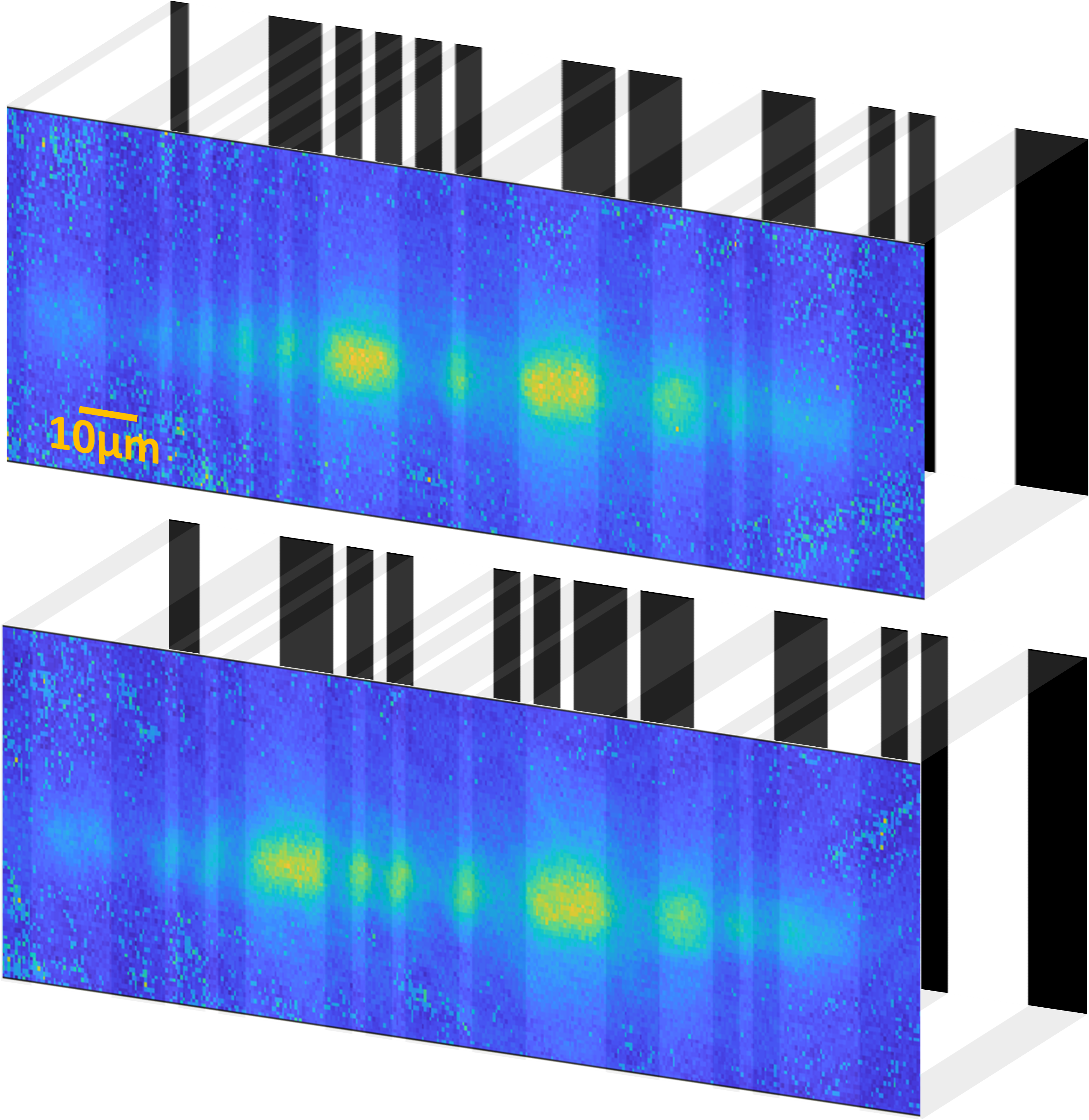}
    \caption{In-situ absorption imaging of our ultracold atoms under the action of two distinct `disordered' potentials. \rcomm{The temperature of the atoms is 20 nK}. The patterns imprinted on the DMD are reported in the back panels, and we indicate with black (white) the regions of the DMD where the micromirrors are set to off (on). Regions where the micromirrors are on correspond to regions with full intensity on the atoms, and therefore to attractive potentials. \rcomm{The depth of the potential is 90 nK}. The atoms re-arrange in such `disordered' potentials, nicely retracing the patterns set on the DMD.}
    \label{figpatterns}
\end{figure}

Using the SLD, with our setup we can achieve 1 mW of power over an area of 207$\times$117 $\mu$m  at the position where the atoms are trapped. To evaluate the corresponding optical potential, we fit the amplified spectrum of Fig. 1b with a double-Gaussian obtaining the power spectrum $G(\lambda)$, with $\lambda$ the wavelength, and then we calculate
\begin{equation}
    U(x,y)=-\frac{\pi c^2}{2}\Gamma\int\left(\frac{2}{\Delta_2\omega_2^3}+\frac{1}{\Delta_1\omega_1^3}\right)J(x,y,\lambda) d\lambda
\end{equation}
with $\Gamma$ the linewidth and $\omega_{1,2}$ the frequencies of the D1 and D2 lines of $^{87}$Rb, $\Delta_{1,2}=\omega_{1,2}-2\pi c/\lambda$ the corresponding detunings, and 
\begin{equation}
    J(x,y,\lambda)=\frac{2P}{\pi w^2}\frac{G(\lambda)}{\int G(\lambda)d\lambda}e^{-2\frac{x^2+y^2}{w^2}},
\end{equation}
where $P$ is the total power at the position of the atoms and $w$ is the waist of the beam obtained with a Gaussian fit of the intensity profile of Fig. 1e. Similarly, we calculate the heating rate as
\begin{equation}
    \Gamma_h(x,y)=\frac{\pi c^2\Gamma^2}{2\hbar}\int \frac{E_r}{k_B}\left(\frac{2}{\Delta_2^2\omega_2^3}+\frac{1}{\Delta_1^2\omega_1^3}\right)J(x,y,\lambda) d\lambda,
\end{equation}
with $E_r=2\pi^2\hbar^2/m\lambda^2$ the recoil energy, where $m$ is the mass of the atoms, and $k_B$ the Boltzmann constant.
With our setup we can obtain a maximum intensity of $2P/(\pi w^2)=$14.3 W$/$cm$^2$, corresponding to a maximum potential depth $U/k_B\simeq90$ nK. This red-detuned potential is therefore sufficiently strong to efficiently manipulate our ultracold sample. The associated maximum heating rate is instead $\simeq$1 nK/s \footnote{\rcomm{For comparison, the heating rate for a laser at 840 nm with the same intensity is $\simeq0.5$ nK/s.}}. \rcomm{ Indeed, we could not observe any measurable heating in the typical time scale of our experiments (hundreds of ms) even when illuminating the atomic cloud with the DMD fully on.}
\begin{figure}
    \centering
    \includegraphics[width=0.48\textwidth]{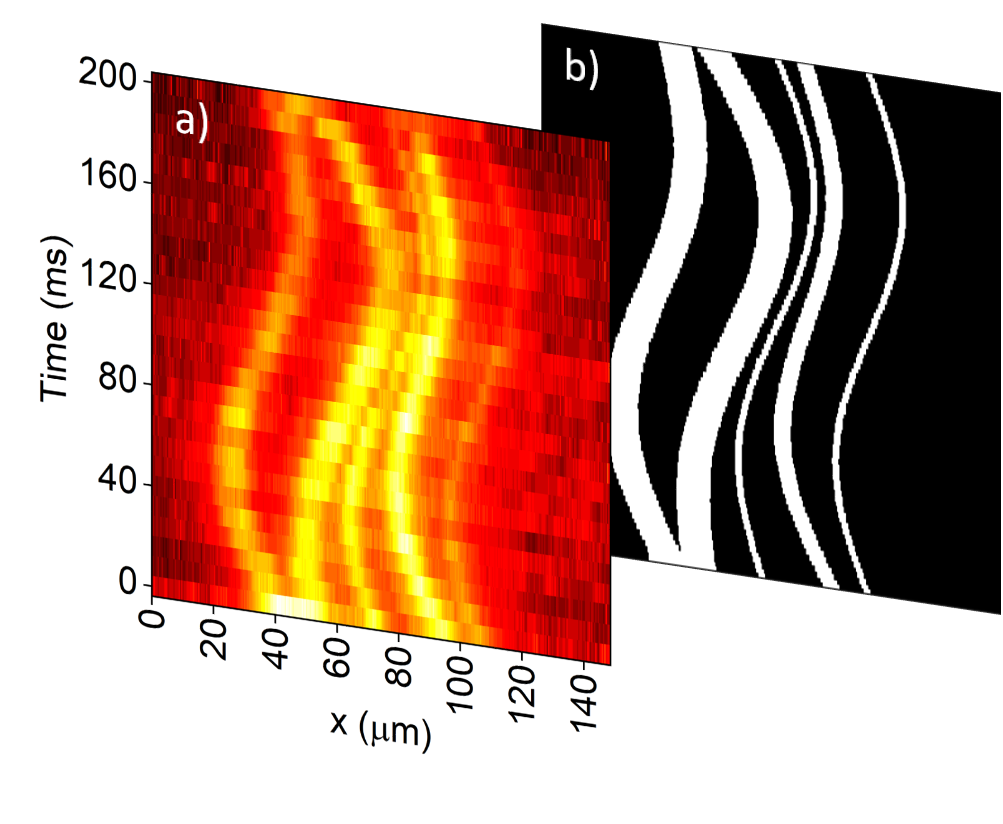}
    \caption{a) Evolution of the atomic column density, integrated along the $y$ direction, as a function of time under the action of the dynamical potential (also integrated along the $y$ direction) displayed in panel b). The time frames are taken every 8 ms. The black (white) regions in b) correspond to micromirrors set off (on) on the DMD.}
    \label{figdynamic}
\end{figure}

To give an example, in Fig. 5 we demonstrate that the combination of arbitrary patterns produced with a DMD and the good quality of the images produced by the SLD can be used to arrange ultracold atoms in non-periodic structures. We have implemented two arbitrary `disordered' 1d potentials by setting unequally spaced gratings on the DMD, as displayed in the back panels. The white vertical lines, corresponding to regions of high intensity, cross with the dipole trap potential creating a series of neatly-separated, sharp squared dimples of different sizes and with different spacing. The dimples are sufficiently deep to trap the atoms, indeed their density profile in absorption nicely retraces the patterns imprinted on the DMD, including the sharp edges. \rcomm{Evaluating the fidelity and contrast directly on the the atomic sample can therefore also be used to characterise the potential.}

We finally demonstrate the implementation of dynamical potentials with SLDs. The possibility to rapidly reconfigure the pattern on the DMD is particularly appealing as it allows us to dynamically and arbitrarily modify the potential on the atoms a fast as one frame every 100 $\mu$s. For dynamical potentials, good image quality is especially desirable as it ensures a good uniformity of the potential across the whole illuminated area, potentially reducing losses of atoms, spurious excitations and heating. In Fig. 5a we report the evolution of the density profile of our ultracold atomic sample under the action of the dynamically varying `disordered' potential of Fig. 5b. For every frame we have integrated the density profiles along the $y$ direction. In this case, the pattern on the DMD was modified every 500 $\mu$s. We observe that the atomic density is dynamically modified following the trajectory of the optical potential. After 200 ms of evolution, we retain $\simeq$75\% of the atoms.  

\section{Conclusion}
In conclusion, we have investigated the use of a SLD in combination with a DMD to produce optical potentials for ultracold atoms. We have compared the ability of a SLD and a laser to produce regular square-wave gratings with different periodicity. We have shown that by using a SLD we consistently obtain images with higher fidelity than and comparable contrast to a laser, resulting in an overall superior quality over the whole range of parameters explored. We have used the patterns generated with the SLD to produce arbitrary 1d disordered potential for our ultracold atoms. This has resulted in patterned atomic densities that nicely retrace the optical potential. We have further demonstrated the implementation of dynamical 1d potentials with our system. Notably, the setup studied here is simple, robust, cost-effective, and, as such, can easily find application in many ultracold atoms experiments worldwide. Additionally, SLDs are available in the near infrared and across most of the visible spectrum, making them suitable for a wide range of possible applications in quantum technology. Our results can open new avenues in the study of 1d disordered systems and transport phenomena in 1d. The extension to two-dimensional patterns in combination with a flat trap could be relevant for applications in atomtronics, phase imprinting and quantum simulations experiments. The further extension to three-dimensional patterns using two or more DMDs has the potential of enabling more pristine control in quantum simulators.

\textit{Acknowledgments.} We acknowledge financial support by the EPSRC (Grant No. EP/R021236/1) and fruitful discussions with the members of the Cold Atoms Group at the University of Birmingham.

\bibliography{bibliography.bib}

\end{document}